\documentclass[twocolumn,preprintnumbers,amsmath,amssymb]{revtex4}
\usepackage{graphicx}
\usepackage{amsmath, amssymb}
\begin{document}
\title{Confinement induces conformational transition of semiflexible polymer rings to figure eight form}
\author{Katja Ostermeir}
\altaffiliation{Authors contributed equally.}
\author{ Karen Alim}
\altaffiliation{Authors contributed equally.}
\author{Erwin Frey}
\affiliation{Arnold Sommerfeld Center for Theoretical Physics and Center for NanoScience, Department of Physics,\\ Ludwig-Maximilians-Universit\"at M\"unchen, Theresienstra\ss e 37, D-80333 M\"unchen, Germany}
\email{frey@lmu.de}
\date{\today}
\begin{abstract}
Employing Monte Carlo simulations of semiflexible polymer rings in weak spherical confinement a conformational transition to figure eight shaped, writhed configurations is discovered and quantified. 
\end{abstract}
\maketitle

The conformation of biopolymers is an important aspect for their functionality. For DNA, transcription and replication are governed by specific binding of proteins, a mechanism strongly connected to polymer configuration \cite{Gowers:2003p8538,vandenBroek:2008p281}. Furthermore, conformational transitions of cytoskeletal filaments represent small engines \cite{Schmid:2004p8586}, an idea that might be transferable to build biomimetic nano-actuators. Both biological processes and technological applications of biopolymers are well-studied in \emph{in vitro} setups. Inevitably and sometimes also desirably accompanied with these experiments is the confinement of polymers, for instance, into channels  \cite{Reisner:2005p2443} or micro-chambers \cite{CosentinoLagomarsino:2007p5633}. Confinement is an effect that also arises ubiquitously in biological systems due to cellular compartments and bacterial or viral envelopes. Indeed, confinement affects polymer conformation and induces conformational transitions as shown by the present work concerning semiflexible polymer rings. As an omnipresent form for DNA \cite{Witz:2008p301} and as a new nano-biomaterial building block \cite{Tang:2001p1647,Sanchez:2010p7457}, semiflexible polymer rings are recently an object of growing interest. Especially biopolymer's resistance against bending on length scales of their persistence length $l_p$, their semiflexibility, turns them into an interesting material, as the degree of overall bending can be tuned by changing their absolute length $L$. The internal structure of polymers can be well characterized by the ``self-crossing number'', the writhe \cite{Panyukov:2001p1656,Norouzi:2008p284} or by correlation functions along the polymer backbone \cite{Morrison:2009p2460}. Thus, we can assess conformational transitions due to confinement.

In this work we investigate the internal structure of semiflexible polymer rings in spherical confinement established by an impenetrable shell. Employing Monte Carlo simulations we compare unconfined polymer rings and polymer rings restricted by different degrees of spherical confinement over the full range of flexibilities. A conformational transition is observed to arise in the semiflexible regime within weak confinement, non-existing in the stiff regime; the mean absolute writhe exhibits an sharp growth to up to two and a half times the unconstrained value. Evaluation of the writhe distribution for different flexibilities reveals from the semiflexible regime onwards a tremendous increase of polymer configurations with writhing numbers specifically centered around $|Wr|=0.8$ within confinement. Finally, the tangent-tangent correlation discloses the conformational transition to figure eight shaped polymer rings due to spherical confinement.

Semiflexible polymers are well described as a concatenated chain of $N$ segments, with tangent vector $\mathbf{t}$,  where the range of the angle between successive bonds is narrowed by the elastic bending energy  $E=Nk_{b}T(l_p/L)\sum_{i=0}^{N}(1-\mathbf{t}_{i}\mathbf{t}_{i+1})$ in the worm-like chain model \cite{KRATKY:1949p2411}. The flexibility $L/l_p$ therein determines the stiffness against bending undulations provoked by thermal energy $k_B T$. A polymer ring is considered stiff, {\it i.e.}, dominated by elastic forces, for flexibilities up to $L/l_p\approx 5$ \cite{Alim:2007p25,Sanchez:2010p7457}, beyond semiflexible behavior smoothly crosses over into the entropic, flexible regime for large $L/l_p$. Polymer conformations are investigated by a Metropolis Monte Carlo simulation, where successive configurations of a closed polygon are generated by crankshaft moves. To collect uncorrelated data only every $10^5$th of successive configurations is considered.  Polymer conformations that violate the spherical confinement are excluded when sampling a set of $10^5$ uncorrelated polymer configurations. The statistical error of these ensembles lies within the ranges of the symbols of all data shown.     

 \begin{figure}[t]
\centering
\includegraphics[width=0.45\textwidth]{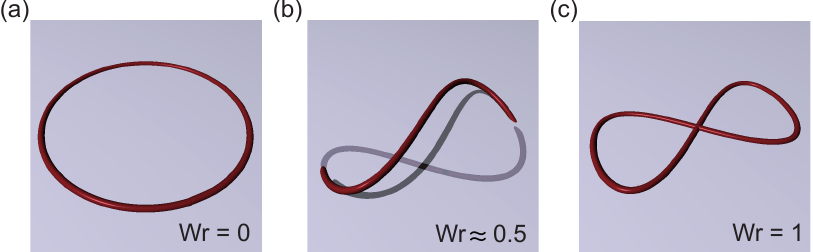}
\caption{Writhe of exemplary polymer ring conformations. (a) When no angular perspective reveals any self-crossings as for a fully symmetric ring the writhe is zero. (b) Writhing occurs when for example an ellipse is twined about itself, here to the degree of $Wr\approx 0.5$. (c) A point of self-intersection increases the writhe by one as in the case of this planar figure eight shaped trajectory. }
\label{fig_cartoon}
\end{figure}
The internal structure of polymers can be assessed by the correlation of two tangent vectors separated a distance $s\in[0,L]$ along the polymer backbone $\langle\mathbf{t}(s)\mathbf{t}(0)\rangle$. This observable provides details about the relative orientation of the whole contour line of a polymer, however, it fails to reflect the position of polymer segments with regard to each other in space.  This aspect is considered by the writhe $Wr$ \cite{FULLER:1971p8445} , which measures the degree of coiling of a polymer by counting the number of crossings of the polymer with its own axis. Projecting a three-dimensional polymer trajectory into a plane defined by a normal vector $\mathbf{n}$ results in a  two-dimensional curve, which may exhibit crossings. Counting these crossing with $\pm1$ according to their handedness and averaging the number of crossings over all angular perspectives given by all possible normal vectors $\mathbf{n}$ defines the writhe $Wr$ of the three-dimensional trajectory. Hence, a two dimensional curve always exhibits integer writhing numbers, {\it i.e.}, $Wr=0$ for a circle and $Wr=1$ for a figure eight shaped trajectory, while three dimensional objects in general are characterized by  a real number as shown in Fig.~\ref{fig_cartoon}. As only the orientation in which a trajectory is traced decides if  the writhe is positive or negative, any writhe distribution is symmetric about the origin with the mean writhe being equal to zero. Insights are therefore gained when measuring the mean absolute writhe $\langle |Wr|\rangle$ of a writhe distribution. To calculate the writhe of polymer configurations generated by Monte Carlo simulations we follow Klenin and Langowski \cite{Klenin:2000p2775}. Originally, the writhe has been employed to characterize the supercoiled state of nicked DNA \cite{Bauer:1980p8490,White:1986p8501},  recently, it has been extended as a measure for the increased complexity of random polygons due to knotting in strong confinement \cite{Arsuaga:2002p7056}. Our work considers non-nicked, {\it i.e.}, zero linking number, polymer rings in the semiflexible regime, where knotting is prevented by high bending energy cost.  In addition, the confinement imposed in our study is very weak: The radius $R$ of the restricting sphere is greater or equal than the contour radius $R_c=L/2\pi$ of the polymer's corresponding rigid ring. Thus, the writhe is expected to reflect only the increase in undulations by a linear growth with flexibility $L/l_p$, as predicted for stiff, unconfined polymer rings \cite{Tobias:2000p1716,Maggs:2001p5613}.

\begin{figure}[t]
\centering
\includegraphics[width=0.45\textwidth]{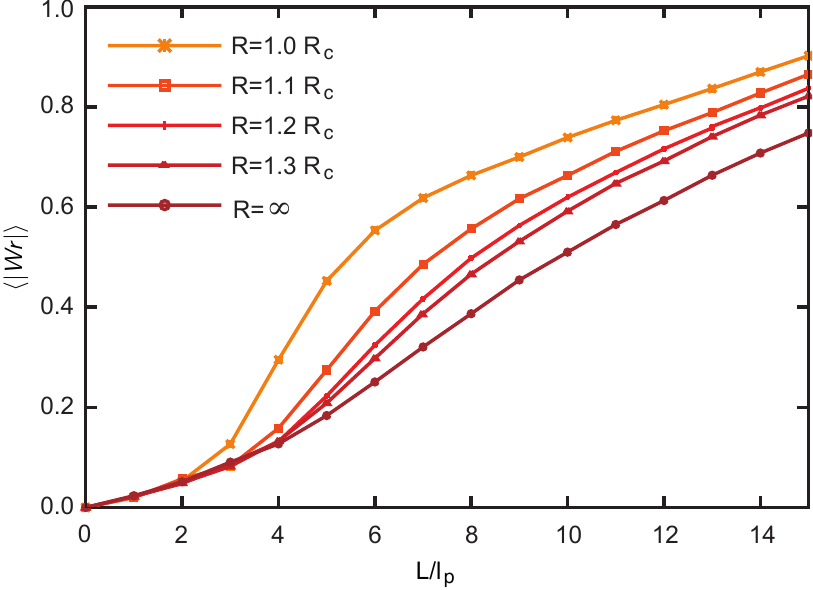}
\caption{Mean absolute writhe versus flexibility $L/l_p$ for unconfined polymer rings $R=\infty$ and polymer rings in weak spherical confinement $R=1.3R_c,\dots,1.0R_c$ of the order of the extent of the polymer's corresponding rigid ring $R_c=L/2\pi$. From flexibilities of about $L/l_p\approx 4$ onwards the confinement induces a sharp increase in absolute writhe proportional to the degree of confinement.}
\label{fig_wr}
\end{figure}
Despite the weak confinement considered, the mean absolute writhe of semiflexible polymer rings displays a strong increase proportional to the degree of confinement for flexibilities $L/l_p>4$ as shown in Fig.~\ref{fig_wr}. The mean writhe grows roughly linearly for stiff, unconfined polymer rings\footnote{
\label{note1}
Expressing the continuous version of the wormlike chain model in Euler angles $\varphi(s),\vartheta(s)$,  $E=l_pk_BT/2\int_0^Lds\{(\partial s\varphi(s))^2\sin^2\vartheta(s)+(\partial s\vartheta(s))^2\}$ with $\partial s=\partial/\partial s$, statistics of stiff, unconfined polymer rings are amenable to calculations by expanding the elastic energy around a rigid ring, {\it i.e.}, $\varphi(s)=2\pi s/L+\delta\varphi(s)$, $\vartheta(s)=\pi/2+\delta\vartheta(s)$. 
Taking the Fourier transform and respecting the periodic boundary conditions then yields, 
$E=k_BT2\pi^2l_p/L\sum_{n=2}^{\infty}\left\{n^2|\delta\varphi^2(n)|+(n^2-1)|\delta\vartheta^2(n)|\right\}$. Applying the equipartition theorem the correlations of the Euler angles are calculated $\langle\delta\varphi(s)\delta\varphi(0)\rangle$, $\langle\delta\varphi(s)\delta\varphi(0)\rangle$. Transferring the reasoning of  Panyukov and Rabin \cite{Panyukov:2001p1656} forecasting $\langle Wr^2\rangle=\sum_{n=2}^Nn^2\{\langle\delta\varphi^2(n)\rangle\langle\delta\vartheta^2(n)\rangle-\langle\delta\varphi(n)\delta\vartheta(n)\rangle^2\}$, the second moment of the writhe fluctuations of a stiff polymer ring are calculated to grow with the square of the flexibility $\langle Wr^2\rangle=\frac{3}{16\pi^4}(\frac{L}{l_p})^2$.
} 
in agreement with previous considerations \cite{Tobias:2000p1716,Maggs:2001p5613}.  Irrespective of the degree of confinement all curves collapse on the unconfined state in the very stiff regime up to $L/l_p\approx 3$. From there on the curves of confined polymer rings start to deviate from the unconstrained case, with strongest confinement rising first. The increase in writhe is very sharp and only saturates on an almost linear growth for higher flexibilities. Towards even higher flexibility we expect the absolute writhe to grow with the square root of polymer length $\sqrt{L}$ as found for the flexible limit of random polygons in spherical confinement  \cite{Micheletti:2006p5130}. The deviation in mean absolute writhe between the different degrees of confinement decreases slightly with growing flexibility.

\begin{figure*}[t]
\centering
\includegraphics[width=0.9\textwidth]{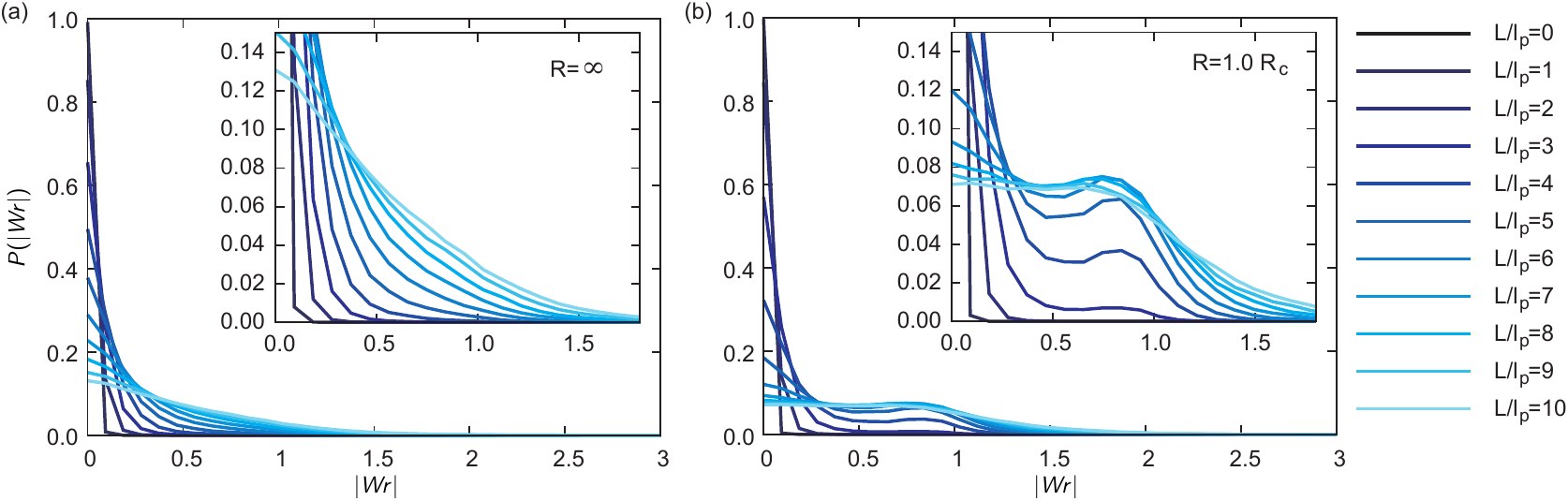}
\caption{Distribution of the absolute writhe for an unconfined polymer ring (a) and a polymer ring within weak spherical confinement of radius $R=R_c$ (b). Displayed are integer flexibilities from $L/l_p=1$ to $L/l_p=10$. The writhe distribution is a monotonically decaying function of the absolute writhe for unconfined semiflexible polymer rings. In contrast, in spherical confinement polymers becomes bimodal by developing in addition to the maximum at $Wr=0$ a well visible local maximum around $|Wr|=0.8$ from $L/l_p\approx 3$ onwards. For sufficiently high flexibilities the two maxima overlap to form a plateau extending up to $|Wr|=0.8$.}
\label{fig_wrdistr}
\end{figure*}
\begin{figure}[b]
\centering
\includegraphics[width=0.45\textwidth]{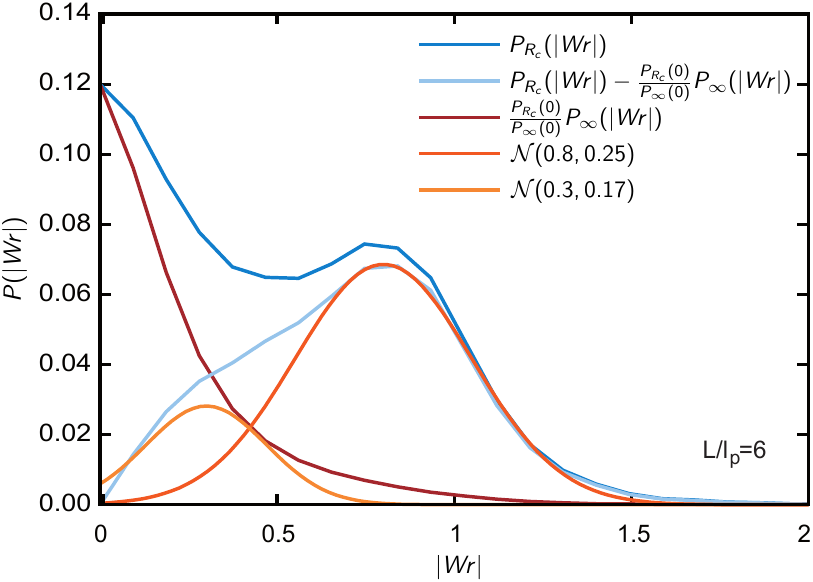}
\caption{Polymer conformations contribution to the writhe distribution of a confined polymer ring $P_{R_c}(|Wr|)$ exemplarily shown for $L/l_p=6$. Subtracting from the confined distribution the decay in writhe from $Wr=0$ as exhibited by the unconfined writhe distribution $P_{\infty}(|Wr|)$ reveals additional states, which can be attributed to two normal distributions $\mathcal{N}(\mu,\sigma^2)$ centered at $|Wr|=0.3$ and $|Wr|=0.8$.}
\label{fig_distr6}
\end{figure}

To understand the sharp increase in mean absolute writhe for confined polymer rings we compare the full distribution of the absolute writhe in both extreme cases considered, unconfined and spherical confinement with radius $R=R_c$ as shown in Fig.~\ref{fig_wrdistr}. The writhe distribution of a free semiflexible polymer ring is monotonically decaying from $Wr=0$ continuously spreading out with increasing flexibility \cite{Tobias:2000p1716}.\\
In contrast, the writhe distribution of a confined polymer ring displays a very different behavior. While the writhe distributions decay from $Wr=0$ in the stiff limit like in the unconfined case, the distributions from $L/l_p\approx 3$ onwards become bimodal exhibiting a second maximum at $|Wr|= 0.8$. This maximum gains statistical weight at the expense of the first at $Wr=0$ as flexibility grows. At sufficiently high flexibilities both maxima have spread out so much that they overlap to form a plateau that extends up to the maximum at $|Wr|=0.8$ before the writhe distribution decays for large absolute writhe. These qualitative observations can also be quantified by extracting the contributing polymer configurations as shown in Fig.~\ref{fig_distr6}. 
Assuming that free $P_{\infty}(|Wr|)$ and confined $P_{R_c}(|Wr|)$ polymer ensembles show the same decay from $Wr=0$, the contributing configurations from this decay can be subtracted from the full absolute writhe distribution of confined polymer rings by $P_{R_c}(|Wr|)-\frac{P_{R_c}(0)}{P_{\infty}(0)}P_{\infty}(|Wr|)$. This discloses the underlying contributing polymer conformations, in particularly a distribution centered around $\langle|Wr|\rangle=0.8$ that is well approximated by a Gaussian $\mathcal{N}(\mu,\sigma^2)$. 
\begin{table}[b]
{\scriptsize
\begin{tabular}{l|c|c|c|c|c|c|c|c}
$L/l_p$ & 3 & 4 & 5 & 6 & 7 & 8 & 9 & 10\\
\hline
\% & 3 & 18 & 34 & 47 & 54 & 55 & 50 & 49\\
$\sigma^2$ & 0.16 & 0.19 & 0.21 & 0.25 & 0.31 & 0.35 & 0.40 & 0.47\\
\end{tabular}}
\caption{Percentage and variance $\sigma^2$ of structural distinct polymer states centered around $\langle|Wr|\rangle=0.8$. Error of 3\% in percentage and 0.03 in variance.}
\label{tab}
\end{table}
Table \ref{tab} displays for different flexibilities the total number of configurations and the variance of the Gaussian distribution that was fitted to extract this information from the data. In addition, at lower absolute writhe Gaussian shaped distributions are observed centered around $\langle|Wr|\rangle=0.17$ for $L/l_p=3$ then shifting to $\langle|Wr|\rangle=0.23$ for $L/l_p=4$ and $\langle|Wr|\rangle=0.3$ for $4<L/l_p\leq6$. Their percentage amounts at most to 20\% and decays strongly for progressing flexibility. The distribution centered around $\langle|Wr|\rangle=0.8$ is fixed in its mean only spreading in variance for increasing flexibilities. Moreover, the absolute percentage of polymer conformations centered around $\langle|Wr|\rangle=0.8$ grows up to over 50\% very slowly decaying in the flexible regime for $L/l_p>8$. Thus, confinement provokes at flexibilities larger than $L/l_p\approx 3$ additional  writhed polymer configuration whose portion amounts up to over 50\% of all states.

As a first step towards an understanding of these observations we consider overall polymer shape. In the limiting case of zero flexibility a polymer ring is a rigid planar ring with zero writhe. For slightly higher flexibility the first bending mode deforms the free polymer ring into a planar, elliptical shape \cite{Alim:2007p25}, whose axes grow and shrink, respectively, with the square root of the flexibility, up to $L/l_p\approx5$. Surely, the writhe of any truly two-dimensional ellipse is zero as well. However, thermal fluctuations do excite small deviations out of the plane such that the actual ensemble of free, stiff polymer rings does exhibit crossings in a small fraction of angular perspectives and, hence, displays a small writhing number. As flexibility grows the writhing number slowly increases.\\
In contrast, even weakly confined polymer rings cannot form the desired elliptical configuration of free polymers as soon as the major axis of the ellipse exceeds the spherical confinement. Instead they buckle into a banana-like ellipse \cite{Ostermeir:2010p8058}. Any symmetrically buckled ellipse again has zero writhe as the mirror plane through the ellipse's apices ensures that any crossing observed from a certain perspective cancels in the summation with a crossing of the opposite sign from the mirror perspective. Hence, very stiff, weakly confined polymer rings only show a small writhe due to undulations about the buckled curve as in the unconfined case. In both cases the absolute writhe grows with flexibility as the undulations increase with $L/l_p$; see Note [23]. Thus, writhe distribution and mean absolute writhe collapse in the stiff regime.

\begin{figure*}[t]
\centering
\includegraphics[width=0.9\textwidth]{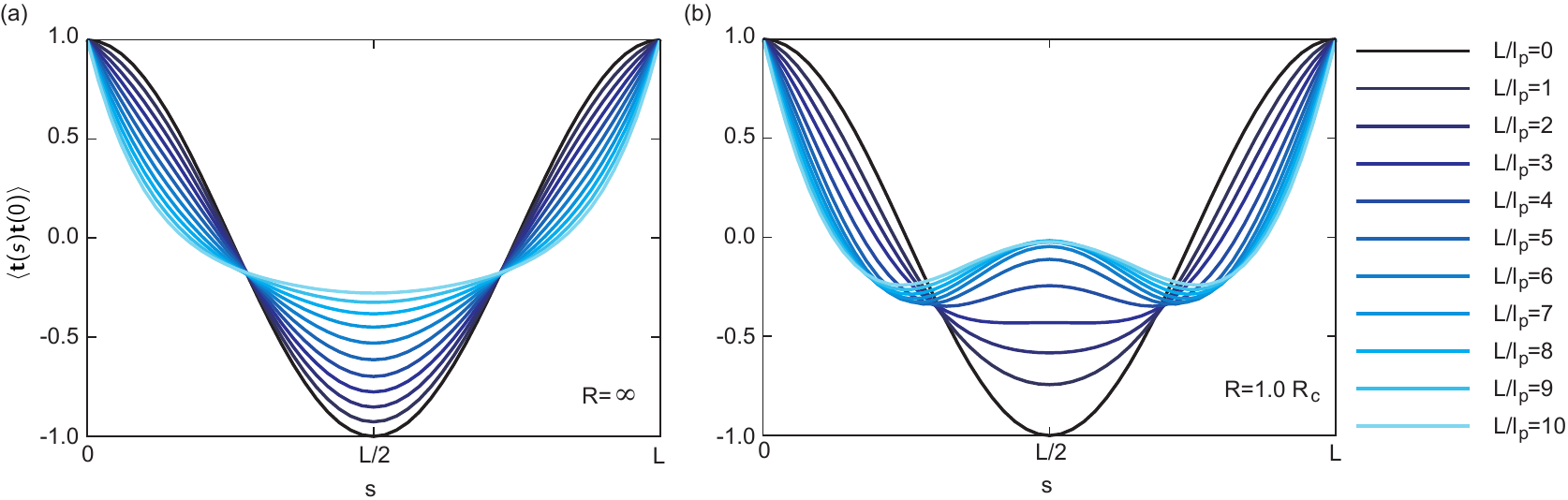}
\caption{Mean tangent-tangent correlation $\langle\mathbf{t}(s)\mathbf{t}(0)\rangle$ along the whole polymer backbone $s\in[0,L]$ for a free (a) and a spherically confined $R=R_c$ (b) semiflexible polymer ring over flexibilities $L/l_p=0,\dots,10$. The correlation of an unconfined polymer ring smoothly crosses over from the tangent-tangent correlation along an ``ellipse'' in the stiff regime to an exponential decay with periodic boundary conditions in the flexible limit. In contrast the correlation of a spherically confined polymer ring displays the correlation of a ``figure eight'' shaped trajectory from $L/l_p>3$ onwards, only slowly saturating for large flexibilities.}
\label{fig_tt}
\end{figure*}
To understand the structural transition which is provoked by the spherical confinement beyond the stiff regime the tangent-tangent correlation $\langle\mathbf{t}(s)\mathbf{t}(0)\rangle$ along the polymer backbone $s\in[0,L]$, is considered in the cases $R=\infty$ and $R=R_c$, see Fig.~\ref{fig_tt}. The symmetry of the correlation function about the point of anti-correlation situated at half the distance along the polymer backbone $s=L/2$ arises due to the topology of a ring. For example, the tangent-tangent correlation of a rigid ring yields $\langle\mathbf{t}(s)\mathbf{t}(0)\rangle=\cos(2\pi s/L)$. In the event of no confinement the elliptically shaped polymers in the stiff regime $L/l_p<5$ display a tangent-tangent correlation that resembles the correlations along an elliptical trajectory in agreement with analytic calculations 
 \footnote{
Employing the calculations of Note [23] the correlation functions of the tangent vector $\langle\mathbf{t}(s)\mathbf{t}(0)\rangle$ are calculated based on its definition in Euler angles $\mathbf{t}(s)=(\sin\vartheta(s)\sin\varphi(s),-\sin\vartheta(s)\cos\varphi(s),\cos\vartheta(s))$. Expanding the tangent product for small angles $\delta\varphi$ and $\delta\vartheta$ and inserting the correlations of the relative Euler angles then yields $\langle\mathbf{t}(s)\mathbf{t}(0)\rangle=\cos(2\pi s/L)[1+\frac{1}{2\pi^2}\frac{L}{l_p}(g(s)-g(0)-f(0))]+\frac{1}{2\pi^2}\frac{L}{l_p}f(s)$ with $g(s)=(\pi-2\pi s/L)^2/4-\pi^2/12-\cos(2\pi s/L)$ and $f(s)=\frac{1}{4}[2+\cos(2\pi s/L)+2\pi(2s-1)\sin(2\pi s/L)]$. Perfect agreement with Monte Carlo simulation data is obtained for flexibilities up to $L/l_p<5$ thus describing the full stiff limit.
}.
In contrast to the correlations of a real ellipse the anti-correlations do not reach down to $\langle\mathbf{t}(L/2)\mathbf{t}(0)\rangle=-1$ due to fluctuations which distort the direction of mirror polymer segments half the backbone distance apart. At higher flexibilities beyond the stiff regime the elliptical character is lost as higher modes crumple up the polymer configuration. This fact is also marked by a change in the initial curvature of the correlation function from convex to concave. With increasing flexibility the topological constraint becomes locally less and less important, thus the correlation function gradually approaches a symmetric exponential decay similar to an open semiflexible polymer. Quite strikingly the tangent-tangent correlations for all flexibilities intersect at $s/L=0.2744\dots,0.7744\dots$ as calculated from the analytic result for the correlation function in the stiff regime presented in Note [24]. This suggests a certain symmetry in the undulations excited in an unconfined semiflexible polymer ring, they all seem to be superpositions of those generated in the stiff limit. This symmetry is, however, broken for spherically confined polymer rings.\\
In the very stiff regime up to $L/l_p\approx3$ the tangent-tangent correlation of a polymer ring in spherical confinement of $R=R_c$ displays an elliptical character, but compared to the unrestricted case the anti-correlation is notably less pronounced. As the confined polymer ring is forced to extend in three dimensional space due to buckling, the probability for deviating directions between mirror segments along the polymer backbone is considerably higher than for an unconstrained planar polymer. Beyond the stiff range the tangent-tangent correlation reveals the internal structure corresponding to the new polymer configurations induced by confinement. For $L/l_p>3$ the correlation function displays decay and increase with twice the frequency as observed for elliptically shaped polymers. In fact, a figure eight shaped trajectory exhibits a correlation function of that frequency given by $\langle\mathbf{t}(s)\mathbf{t}(0)\rangle=\cos(4\pi s/L)$. Indeed an elliptical trajectory that is twined about its longest axis could account for both the correlation function and a writhing number around $\langle|Wr|\rangle=0.8$. The states of smaller writhe are again hidden in the full distribution due to their small percentage in number. The figure eight correlation becomes most pronounced around $L/l_p=7$, when the percentage of $\langle|Wr|\rangle=0.8$ is largest. For higher flexibilities the function smoothes out, as the distribution of states broadens in accordance with observations from the writhe distribution.

Altogether our observations confirm that weak spherical confinement imposed by an impenetrable shell induces a conformational transition to polymer rings above a certain flexibility. Very stiff polymer rings below $L/l_p\approx 3$ exhibit very symmetric conformations whose mean trajectory obeys zero linking number. Only small undulations around this mean trajectory yield a finite mean writhe increasing identically with flexibility for weakly confined and unrestricted polymer rings. Differences in internal structure are only visible by the tangent-tangent correlation reflecting the planar configuration of unconfined and buckled three-dimensional state of confined polymers.\\
Increasing flexibility further beyond this stiff regime induces more and more undulations. For a free polymer ring these lead to more and more crossings in projection planes broadening the writhe distribution. Crumpled configurations do not exhibit the symmetry of an elliptical trajectory any more, which in the stiff limit provoked cancelations in the sum over crossings, hence yielding smaller writhing numbers. Thus, beyond the stiff regime the mean absolute writhe shows a steeper increase with flexibility. In confinement, however, it is not only the increase in undulations that raises the mean writhe with flexibility.  In addition, there is a qualitative change in the writhe distribution. From $L/l_p\approx3$ onwards the writhe distribution becomes bimodal as polymer configurations with a writhing number distinctly distributed around $\langle|Wr|\rangle=0.8$ start to develop. Also the tangent-tangent correlation shows from this threshold onwards the characteristics of a figure eight trajectory. We, therefore, deduce that confinement induces a conformational change to figure eight shaped polymers with $\langle|Wr|\rangle=0.8$. Additional polymer states centered around $\langle|Wr|\rangle=0.17,0.23,0.3$ are few in number and only transiently occur between $3\leq L/l_p\leq6$.  Configurations centered around $\langle|Wr|\rangle=0.8$ grow strongly in percentage with flexibility and only slowly decay from $L/l_p>8$ onwards. In fact, the first bending mode induces an elliptical polymer form that has to buckle transversely within spherical confinement. Thus, it seems plausible that an increase in flexibility and, hence, in ellipse eccentricity but also in possible curvature results in a polymer conformation, where the free energy to bend a polymer into a writhed state is equal or less than the free energy of stronger transverse bending. Therefore, a polymer ring chooses to intertwine with itself as an alternative way to fit inside a sphere.  This picture is also in agreement with the subsequent and less pronounced increase in mean absolute writhe for larger cavities. In larger confinement the energy for buckling becomes comparable to the bending energy paid for writhing at higher flexibility and also to less extend. It is, however, quite remarkable that the structural transition selects specific writhing numbers. Already the tangent-tangent correlation of free polymer rings suggests a selection of bending modes, which might be extended to a selection of ``writhing modes" within confinement. Surely regarding the specific writhing number the exact geometry and strength of confinement enters.  \\
In summary, our work discloses the occurrence and kind of conformational transitions in semiflexible polymers due to an impenetrable shell. The probability curve and the absolute quantity of the restructured state for a given flexibility is accessible from the writhe distribution presented. This turns the conformational transition due to confinement a predictable event to be employed in \emph{in vitro} investigations. Thus, polymer conformations influence on gene regulation or controlled dynamics of conformational transitions due to administered changes in confinement become accessible opening up new perspective both concerning the study of biological process as well as the invention of biomimetic devices. 

The financial support of the Deutsche Forschungsgemeinschaft through SFB 863 and of the German Excellence Initiative via the program ``Nanosystems Initiative Munich (NIM)'' is acknowledged. K.A.~also acknowledges funding by the Studienstiftung des deutschen Volkes.

\end{document}